\documentclass[showpacs,twocolumn,prl,superscriptaddress,notitlepage]{revtex4-2}
\pdfoutput=1
\usepackage{qcircuit}
\usepackage[dvips]{graphicx}
\usepackage{amsmath,amssymb,amsthm,mathrsfs,amsfonts,dsfont,mathtools}
\usepackage{subfigure, epsfig}
\usepackage{braket}
\usepackage{bm}
\usepackage{enumerate}
\usepackage[dvipsnames,svgnames]{xcolor}
\usepackage[colorlinks=true, allcolors=MidnightBlue, linktoc = all]{hyperref}
\usepackage{newtxtext,newtxmath}
\usepackage{stackrel}
\usepackage{changes}

\newtheorem{theorem}{Theorem}

\newcommand{\opt}{\mathrm{}}

\newcommand{\mc}[1]{\mathcal{#1}}

\DeclareMathOperator{\tr}{Tr}

\newcommand{\id}{I}

\newcommand{\bal}{\begin{equation}\begin{aligned}}
\newcommand{\eal}{\end{aligned}\end{equation}}
\newcommand{\sbar}{\;\rule{0pt}{9.5pt}\right|\;}
\newcommand{\lset}{\left\{\left.}
\newcommand{\rset}{\right\}}

\newcommand{\mF}{\mathcal{F}}

\newcommand{\mO}{\mathcal{O}}

\usepackage{xr}
\makeatletter
\newcommand*{\addFileDependency}[1]{
  \typeout{(#1)}
  \@addtofilelist{#1}
  \IfFileExists{#1}{}{\typeout{No file #1.}}
}
\makeatother

\begin{document}
\title{Virtual Quantum Resource Distillation}
\begin{abstract}
Distillation, or purification, is central to the practical use of quantum resources in noisy settings often encountered in quantum communication and computation. Conventionally, distillation requires using some restricted `free' operations to convert a noisy state into one that approximates a desired pure state. Here, we propose to relax this setting by only requiring the approximation of the measurement statistics of a target pure state{, which allows for additional classical postprocessing of the measurement outcomes}. We show that this extended scenario, which we call \emph{virtual resource distillation}, provides considerable advantages over standard notions of distillation, allowing for the purification of noisy states from which no resources can be distilled conventionally. We show that general states can be virtually distilled with a cost (measurement overhead) that is inversely proportional to the amount of existing resource, and we develop methods to efficiently estimate such cost via convex and semidefinite programming, giving several computable bounds. We consider applications to 
coherence, entanglement, and magic distillation, and an explicit example in quantum teleportation (distributed quantum computing). 
This work opens a new avenue for investigating generalized ways to manipulate quantum resources.
\end{abstract}

\author{Xiao Yuan}
\email{xiaoyuan@pku.edu.cn}
\affiliation{Center on Frontiers of Computing Studies, Peking University, Beijing 100871, China}
\affiliation{School of Computer Science, Peking University, Beijing 100871, China}

\author{Bartosz Regula}
\email{bartosz.regula@gmail.com}
\affiliation{Mathematical Quantum Information RIKEN Hakubi Research Team, RIKEN Cluster for Pioneering Research (CPR) and RIKEN Center for Quantum Computing (RQC), Wako, Saitama 351-0198, Japan}
\affiliation{Department of Physics, Graduate School of Science, The University of Tokyo, Bunkyo-ku, Tokyo 113-0033, Japan}

\author{Ryuji Takagi}
\email{ryuji.takagi@phys.c.u-tokyo.ac.jp}
\affiliation{Department of Basic Science, The University of Tokyo, Tokyo 153-8902, Japan}
\affiliation{Nanyang Quantum Hub, School of Physical and Mathematical Sciences, Nanyang Technological University, 637371, Singapore}

\author{Mile Gu}
\email{mgu@quantumcomplexity.org}
\affiliation{Nanyang Quantum Hub, School of Physical and Mathematical Sciences, Nanyang Technological University, 637371, Singapore}
\affiliation{Centre for Quantum Technologies, National University of Singapore, 3 Science Drive 2, 117543, Singapore}
\affiliation{CNRS-UNS-NUS-NTU International Joint Research Unit, UMI 3654, Singapore 117543, Singapore}

\maketitle

One important aspect in the manipulation of any valuable physical resource is distillation~---~a means to extract this resource in optimal, purified form from some crude, noisy source. 
This finds use in quantum information processing, where various quantum resources~\cite{horodecki2013quantumness,coecke_2016,chitambar2018quantum} --- entanglement~\cite{PhysRevA.53.2046,PhysRevLett.78.2275,RevModPhys.81.865}, coherence~\cite{aberg2006quantifying,Baumgratz14,RevModPhys.89.041003}, or magic~\cite{veitch2012negative,veitch2014resource,PhysRevLett.118.090501}, among many others --- have gained interest for their role in empowering various quantum information protocols~\cite{horodecki2013quantumness,coecke_2016,chitambar2018quantum}. However, the inevitable imperfections that permeate near-term quantum technologies mean that noiseless resources may not be readily available. 
Thus, distillation of such resources is critical for developing practical quantum computation and communication schemes, allowing for systematic means to obtain ideal resources from ones distorted by noise.

Powerful theoretical results have been obtained for both one-shot and asymptotic resource distillation~\cite{bennett_1996-1,kent_1998,rains2001semidefinite,brandao2008entanglement,Brandao2011oneshot,zhao2018oneshot,regula2018one,2019arXiv190405840L,Regula2020benchmarking,PhysRevLett.125.060405,Vijayan2020simple,Regula2021fundamental,Fang2020no-go,Regula2021oneshot,Takagi2021oneshot,regula_2022}. Yet, especially in the one-shot scenario, strong limitations exist that prohibit resource distillation even from highly resourceful states, either demanding many copies of the resource state to enable a successful conversion, or incurring large errors in the process~\cite{PhysRevLett.125.060405,Regula2021fundamental,Fang2020no-go,regula_2022}. While those results are fundamental no-go theorems based on the laws of quantum mechanics, could there nevertheless be means to side-step them by somewhat bending the rules?

\begin{figure*}[t]
\includegraphics[width=\textwidth]{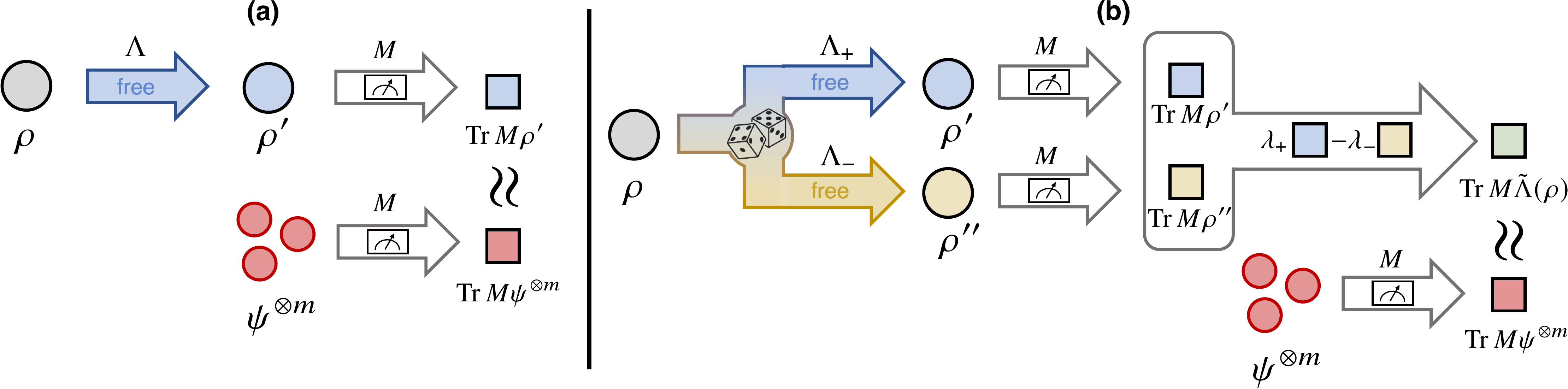}
    \caption{
    Two different approaches to resource distillation. (a) Conventional resource distillation employs a free operation $\Lambda$ to map $\rho$ into a state that approximates the target state $\psi^{\otimes m}$.
    (b) Virtual distillation approximates the measurement outcomes of $\psi^{\otimes m}$ by using the virtual operation $\tilde \Lambda= \lambda_+ \Lambda_+ - \lambda_- \Lambda_-$, a linear combination of free operations.
    }
    \label{fig:fig1}
\end{figure*}

Here, we give an affirmative answer to this question by proposing a new paradigm of \emph{virtual resource distillation}. Since the output of many quantum information protocols is ultimately purely classical, it often does not matter whether a particular ideal resource state (e.g.~a Bell state) was actually synthesized, provided we can reproduce the same expected outcome statistics. {Therefore, we focus on distilling a `virtual' target resource state, in the sense that any operation followed by a measurement of the target resource can be approximated to a desired accuracy.}
This combines the power of both classical data processing and resource manipulation, performing better than either of these two approaches on their own.
We show that virtual distillation enables us to effectively increase the distillation efficiency {at an increased} cost in the measurement samples. 
We study properties of the distillation overhead and show how it can be tightly bounded using resource monotones and semidefinite programs. 
We show that various limitations on distillation schemes can be circumvented by employing our framework, enabling distillation even from states from which no resource can be extracted conventionally. 
We give examples in the virtual distillation of coherence, entanglement, and magic, and discuss applications.

Here we focus on the case where the manipulated objects are quantum states, and we leave the broader framework and complete technical proofs to the companion paper~\cite{sm}.

\vspace{0.2cm}
\textbf{\emph{Background}.\,---}
A resource theory of states consists of the set $\mc F$ of free states and the set $\mc O$ of free operations. {A weak and undemanding assumption about the free operations $\Lambda \in \mc O$ is the so-called `golden rule', which states that} $\Lambda(\sigma)\in \mc F$ 
for all $\sigma\in \mc F$. The set of all operations that satisfy the golden rule, called resource non-generating operations, is then the largest consistent set of free operations.
A resource monotone $\mu$ is a function that measures the amount of resource a state possesses, satisfying the monotonicity requirement $\mu(\rho)\ge \mu(\Lambda(\rho)),\,\forall \Lambda\in \mc O$.

In most resource theories, we can define an optimal unit pure resource state, denoted as $\psi_{\opt}$. Distillation then captures the task of synthesising $\psi_{\opt}$ from an imperfect state $\rho$. The one-shot resource distillation rate 
\begin{equation}\label{eq:one-shotdistilldef}
	\begin{aligned}
		D^\varepsilon(\rho) &= \sup_{\Lambda\in \mc O}\big\{m:\frac{1}{2}\big\|\Lambda(\rho)-\psi^{\otimes m}_{\opt}\big\|_1\le \varepsilon\big\}
			\end{aligned}
\end{equation}
then defines number of optimal states we can synthesize with $\rho$ at allowable error $\varepsilon\in[0,1)$. Here $\|A\|_1 = \tr\big[\sqrt{A^\dag A}\big]$ is the trace norm.

\vspace{0.2cm}
\textbf{\emph{Virtual resource distillation}.\,---} In many settings, {distillation is followed by a processing of the state} $\psi_{\opt}$ 
to produce {an expectation value} $x$ {of some observable}. Thus $\psi_{\opt}$ does not strictly need to actually exist: 
a simulation of $\psi_{\opt}$ that enables accurate retrieval of $x$ is sufficient.
{We will thus assume that the state is ultimately measured. This contrasts with conventional distillation, where $\psi_{\opt}$ is always synthesized physically.}

Formally, consider a task that involves applying some operation $\mc N$ on the resource state (possibly together with other states), and measuring the resulting state in some observable $M$ to retrieve $x$. {In order for a distillation protocol $\Lambda$ to be successful, we thus require that measuring $\mc N \circ \Lambda(\rho)$ approximates the measurement outcomes of $\mc N \left(\psi^{\otimes m}\right)$ for any choice of a channel $\mc N$ and measurement $M$. Since applying a channel $\mc N$ cannot make the error any larger, our requirement for distillation is in fact equivalent to the statement that}
    $\left| \tr M \Lambda(\rho) - \tr M \psi^{\otimes m}_{\opt}\right| \le \varepsilon$
for any Hermitian operator $M$ satisfying $0\leq M\leq I$. 
This condition is {the same as} the one {for conventional distillation in} Eq.~\eqref{eq:one-shotdistilldef}, {so we have gained no advantage.}
However, since $ \tr M \Lambda(\rho)$ is a classical result, we can further apply classical post-processing with different distillation operations.  Specifically, we can consider a linear combination of the classical results $\sum_j\lambda_j \tr \left(M  \Lambda_j(\rho)\right) = \tr \left(M \sum_j \lambda_j \Lambda_j(\rho)\right)$ using different choices of $\{\Lambda_j\}\subseteq \mc O$ and real coefficients $\lambda_j$ satisfying $\sum_j \lambda_j=1$. Grouping free operations with the same sign together, we have
    $\Big| \tr \left[M \big(\lambda_+\Lambda_+(\rho)-\lambda_-\Lambda_-(\rho)\big)\right] - \tr M \psi^{\otimes m}_{\opt}\Big| \le \varepsilon$,
where $\lambda_\pm= \sum_{j: {\rm sign}(j) = \pm1 }\lambda_j\ge 0$, $\lambda_+-\lambda_-=1$, and $\Lambda_\pm = \frac{1}{\lambda_\pm}\sum_{j: {\rm sign}(j) = \pm 1} \lambda_j\Lambda_j$.
This is equivalent to  the virtual distillation condition
\begin{equation}\label{eq:}
    \frac{1}{2}\big\|\tilde \Lambda(\rho) -\psi^{\otimes m}_{\opt}\big\|_1\le \varepsilon.
\end{equation}
where we define $\tilde \Lambda= \lambda_+\Lambda_+ - \lambda_-\Lambda_-$ to be a virtual operation, see also Fig.~\ref{fig:fig1}. 
We note that the condition is independent of the operation $\mc N$ or measurement $M$. 

In practice, we can effectively implement $\tilde \Lambda$ by following a Monte Carlo--based approach that often finds use in the simulation of quantum circuits~\cite{PhysRevLett.115.070501,PhysRevLett.118.090501,seddon_2020,PRXQuantum.2.040361,PhysRevLett.125.150504,Brenner2023optimal}, quantum error mitigation~\cite{Li2017,doi:10.7566/JPSJ.90.032001,PhysRevLett.119.180509}, and the implementation of unphysical processes~\cite{Buscemi2013twopoint,Jiang2021physical,Regula2021operational}. The basic idea is to notice that for any $M$, we have
$\mc \tr M\tilde \Lambda(\rho) = C\big[ {\rm sign(\Lambda_+)}p_+\tr M \Lambda_+(\rho) + {\rm sign(\Lambda_-)} p_-\tr M \Lambda_-(\rho) \big]$.  Therefore, we can obtain $\mc \tr M\tilde \Lambda(\rho)$ by
randomly applying $\Lambda_\pm$ with probability $p_{\pm} =\lambda_\pm/(\lambda_++\lambda_-)$ and multiply each classical outcome by $C{\rm sign(\Lambda_\pm)}=\pm1$.
Here, $C\coloneqq \lambda_++\lambda_-\geq 1$ contributes to a larger variance of the outcome distribution. 
This essentially increases the number of samples by a factor of $C^2$ compared to the case of conventional distillation where resource states $\psi^{\otimes m}$ are available~\footnote{Specifically, according to the Hoeffding inequality~\cite{Hoeffding1963probability}, $\mc O(\log(1/\delta)/\beta^2)$ and
$\mc O(C^2\log(1/\delta)/\beta^2)$ number of samples can estimate $\tr[M\psi^{\otimes m}]$ and $ \tr[M\tilde \Lambda(\rho)]$ to achieve an accuracy $\beta\ge0$ with a failure probability less than $\delta\ge0$.}. 
This means that the effective number of $\psi$ virtually obtained as $\tilde\Lambda(\rho)$ is reduced by a factor of $1/C^2$ for the purpose of estimating the expectation value of an observable with the desired accuracy.

This observation motivates us to define the virtual resource distillation rate as 
\begin{equation}\label{eq:virtualratemax}
	\begin{aligned}
		V^\varepsilon(\rho) &= \sup_{m}\frac{m}{C^\varepsilon(\rho, m)^2},
			\end{aligned}
\end{equation}
with the {overhead} $C^\varepsilon(\rho, m)$ of virtual operations 
\begin{equation}\label{eq:virtualcost}
	\begin{aligned}
		C^\varepsilon(\rho, m) &= \inf_{
		\underset{\Lambda_{\pm}\in \mc O, \lambda_{\pm}\ge 0}{\overset{{\tilde\Lambda=\lambda_+\Lambda_+ - \lambda_-\Lambda_-}}{\lambda_+-\lambda_-=1}}}
		\big\{\lambda_+ + \lambda_-:\frac{1}{2}\big\|\tilde\Lambda(\rho)-\psi^{\otimes m}_{\opt}\big\|_1\le \varepsilon\big\}.
			\end{aligned}
\end{equation}
The virtual distillation rate $V^\varepsilon$ generalizes the conventional one-shot distillation rate $D^\varepsilon$, recovering the latter when optimization in \eqref{eq:virtualcost} is restricted to the case of $\lambda_-=0$. 
This immediately implies that   $ D^\varepsilon(\rho) \leq V^\varepsilon(\rho)$. 
Since $V^\varepsilon(\rho)$ is fully determined by $C^\varepsilon(\rho, m)$, we focus on the estimation of $C^\varepsilon(\rho, m)$ in the following.

{We note here two major differences between the settings of virtual and conventional distillation.
First, our approach is to perform distillation by repeatedly applying free operations to a noisy resource state to obtain a sufficient number of samples. This is very different from conventional distillation protocols used in practice, which work by employing \emph{joint} operations on multiple copies of a state~\cite{bennett_1996-1,bravyi_2005}. Coherent many-copy manipulation through joint operations is experimentally much more difficult to realize, often making resource distillation challenging or inefficient in practice. Virtual distillation overcomes this by only requiring single-copy operations.
Our approach can also be compared with \emph{probabilistic} one-shot distillation protocols in the conventional case, which also repeatedly realize single-copy operations; we address this in more detail in~\cite{sm}. We will shortly see that virtual distillation does not suffer from limitations that constrain conventional (including probabilistic) distillation in equivalent settings.

A second, perhaps counterintuitive difference is that virtual distillation allows for the distillation from free (resourceless) states at a non-zero rate. This becomes possible as we relaxed the distillation requirement to only approximate the outcome statistics:
{for a free input state, our method essentially reduces to classical data processing,} and simply reflects that the cost of classically simulating the target state $\psi$ using free states $\mF$ is not infinite. 
}\\

\begin{figure*}[t]
    \centering
    \includegraphics[width =\linewidth]{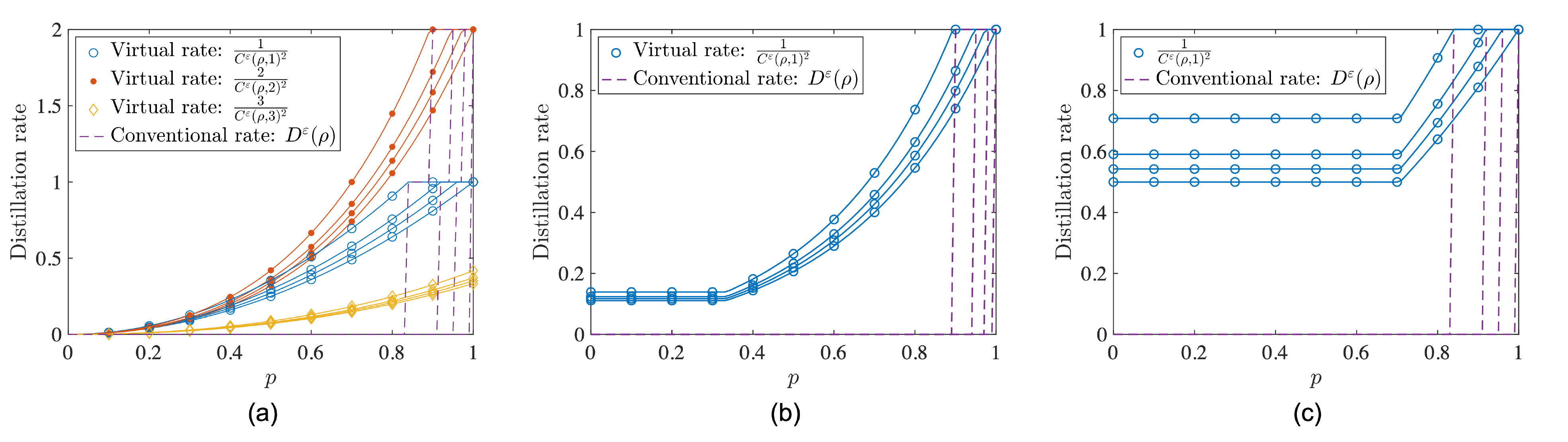}
\caption{Comparison between conventional and virtual one-shot resource distillation. We consider four different smoothing parameters $\varepsilon=\{0,0.02,0.04,0.08\}$ corresponding to each of the four lines (with the same color or dot type) from bottom to top.
    (a) Distillation from the four-dimensional coherent state $\rho_p=p\psi_4 +(1-p) I_4/4$ under MIO/DIO. Here  $\ket{\psi_4}=(\ket{0}+\ket{1}+\ket{2}+\ket{3})/2$ and $I_4$ is the four-dimensional identity matrix.  We also consider virtual distillation with different $m$ as in Eq.~\eqref{eq:virtualratemax} and find that $m=2$ outperforms the others for most cases (except for very small $p$ where $m=1$ performs the best). (b) Distillation of the two-qubit entangled state $\rho_p=p\psi + (1-p)I_4/4$ under SEPP operations. Here $\ket{\psi}=(\ket{00}+\ket{11})/\sqrt{2})$, and the virtual distillation rate is achieved with LOCC already.  
    (c) Distillation of the noisy magic state $\rho_p=pT + (1-p)I_2/2$ under stabilizer operations. The magic state $T$ is defined as $T = \frac12 \left(I+[X+Y]/\sqrt{2}\right)$, where $X$ and $Y$ are Pauli matrices.
   The virtual distillation rates for (b) and (c) are obtained with $m=1$ as in Eq.~\eqref{eq:virtualratemax}. 
   We note that virtual distillation also outperforms probabilistic distillation for these examples. {See~\cite{sm} for details.}%
   }
    \label{fig:results}
\end{figure*}

\textbf{\emph{Estimation of $C^\varepsilon(\rho, m)$}.\,---}
We introduce bounds on $C^\varepsilon(\rho, m)$ in general resource theories, which 
 rely on {two ingredients. First, we introduce the optimization problem:}
\begin{equation}\label{eq:zetadef}
\begin{aligned}
	\zeta^s_\varepsilon(\rho, k) \coloneqq \text{minimize} &&& \mu_+ + \mu_- \;\\
	\textrm{s.~t.}&&& 0 \leq Q_+ \!\leq \mu_+ \id,\;\, 0 \leq Q_- \!\leq \mu_- \id,\\
	&&& \tr Q_+ \sigma \!\leq\! \frac{\mu_+}{k}, \; \tr Q_- \sigma \!\leq\! \frac{\mu_-}{k} \; \forall \sigma \in {\mc F}\!,\\
	&&& \mu_+ - \mu_- \!= 1,\;\,
	\tr \rho (Q_+ - Q_-) \geq 1\!-\! \varepsilon ,
\end{aligned}
\end{equation}
where $k$ is some parameter to be fixed. 
We also define $\zeta^{g}_\varepsilon(\rho, k)$ to be the same optimization except that the inequality constraints in the second line become equality constraints: $\tr Q_+ \sigma = {\mu_+}/{k}, \; \tr Q_- \sigma = {\mu_-}/{k} \; \forall \sigma \in {\mc F}$. 
{%
The second ingredient are} three different resource measures for the target pure resource state $\psi$~---~the generalized robustness $R^g_{\mc F} (\rho) \coloneqq \inf \lset \lambda \sbar \frac{\rho + \lambda \omega}{1+\lambda} \in \mc F,\; \omega \in \mc D \rset$, the standard robustness $R^s_{\mc F} (\rho) \coloneqq \inf \lset \lambda \sbar \frac{\rho + \lambda \sigma}{1+\lambda} \in \mc F,\; \sigma \in \mc F \rset$, and the resource fidelity $F_{\mc F} (\rho) \coloneqq \max_{\sigma \in \mc F}  \left(\tr\sqrt{\rho^{1/2}\sigma \rho^{1/2}}\right)^2$.
Here $\mc D$ is the set of all density matrices. 
{Our first main result is as follows.}

\begin{theorem}\label{prop:general_bounds_cost} {(Theorem 1 in the companion paper~\cite{sm})}
Consider a convex resource theory and a target pure resource state $\psi$. 
Let $\mc O$ be the class of resource non-generating operations. If $R^s_{\mc F}(\psi) < \infty$, then
   $\zeta^s_\varepsilon\left(\rho, F_{\mc F}(\psi^{\otimes m})^{-1}\right)\le C^\varepsilon(\rho,m) \leq  \zeta^s_\varepsilon\left(\rho, R^s_{\mc F} (\psi^{\otimes m}) + 1 \right)$.
Furthermore, if it holds that $\braket{\psi|\sigma|\psi}$ is constant for all $\sigma \in \mc F$, then 
 $\zeta^g_\varepsilon\left(\rho, F_{\mc F}(\psi^{\otimes m})^{-1}\right)\le C^\varepsilon(\rho,m) \leq \zeta^g_\varepsilon\left(\rho, R^g_{\mc F}(\psi^{\otimes m}) + 1 \right)$.
\end{theorem}
\noindent The crucial property of the bounds is that whenever $R^s_{\mc F} (\psi^{\otimes m}) + 1 = F_{\mc F}(\psi^{\otimes m})^{-1}$ --- which is true in resource theories such as bi- and multi-partite entanglement or multi-level quantum coherence --- or if $R^g_{\mc F}(\psi^{\otimes m}) + 1 = F_{\mc F}(\psi^{\otimes m})^{-1}$ and the overlap $\braket{\psi|\sigma|\psi}$ is constant~---~which is true in resource theories such as coherence or athermality~---~then the upper and lower bounds coincide, yielding an exact expression for the overhead $C^\varepsilon(\rho,m)$, i.e., 
\begin{equation}
C^\varepsilon(\rho,m) = \zeta^{s/g}_\varepsilon\left(\rho, F_{\mc F}(\psi^{\otimes m})^{-1}\right).
\end{equation}

\noindent{Recall that the problems $\zeta_\varepsilon^{s/g}$ are convex optimization problems and are often (e.g.\ for coherence, magic states, or non-positive partial transpose) semidefinite programs (SDPs), which are efficiently computable in the dimension of the state space~\cite{sdp_1996}, allowing for an exact evaluation of $C^\varepsilon$ in relevant cases. Importantly, they remove the need to optimize over all operations $\Lambda \in \mO$, which may be a significantly more difficult problem.}

We note that the question of whether there exist suitable states for which the two bounds coincide is, in general, highly dependent on the resource theory in question~\cite{sm}.

Next, we show that the overhead is not only computable numerically, but in fact an exact expression for it can be obtained in terms of another resource monotone, $f_{\mathcal{O}}(\rho,m)$, which measures the maximum overlap between $\Lambda(\rho)$ and  $\psi^{\otimes m}$ as 
\begin{equation}
    f_{\mathcal{O}}(\rho,m)\coloneqq \max_{\Lambda\in\mathcal{O}}\tr[\Lambda(\rho)\psi^{\otimes m}].
\end{equation}
This is useful because $f_\mO$ has been computed exactly in many relevant cases~\cite{rains2001semidefinite,Brandao2010reversible,regula2018one,regula_2019-2,zhao2018oneshotdistill,Regula2020benchmarking}.
Such an equivalence will be possible whenever there exists a free ``generalized twirling'' operation~\cite{Takagi2021oneshot} $\mathcal{T}\in\mathcal{O}$ of the form 
 $\mathcal{T}(\rho)=\tr[\psi^{\otimes m}\rho]\,\psi^{\otimes m} + \tr[(I-\psi^{\otimes m})\rho]\,\sigma^\star$
for some $\sigma^\star\in\mathcal{F}$, which is true for many resource theories of practical interest, such as entanglement and magic theory for specific target states. 
\begin{theorem}\label{thm:theoremtwirling}
{(Theorem 2 in the companion paper~\cite{sm})}
Suppose a free generalized twirling operation exists. Then
\bal\label{Eq:costtwirling}
 C^\varepsilon(\rho,m)=\max\left\{\frac{2(1-\varepsilon)}{f_{\mathcal{O}}(\rho,m)}-1,1\right\}.
\eal
\end{theorem}

\noindent Ref.~\cite{Takagi2021oneshot} showed that the existence of a suitable resource non-generating twirling operation is guaranteed when $F_{\mathcal{F}}(\psi^{\otimes m})^{-1}$ equals $1+R_{\mathcal{F}}^s(\psi^{\otimes m})$. We thus have an alternative characterization of $C^\varepsilon(\rho,m)$ via the resource monotone $f_{\mc O}(\rho, m)$: whenever the condition $F_{\mathcal{F}}(\psi^{\otimes m})^{-1}=1+R_{\mathcal{F}}^s(\psi^{\otimes m})$ is satisfied, $C^\varepsilon(\rho,m)$ under resource non-generating maps is given exactly by Eq.~\eqref{Eq:costtwirling}.

{%
We note that while Theorem~\ref{prop:general_bounds_cost} provides an exact characterization of the virtual distillation overhead without the need for an explicit optimization over the allowed free operations, Theorem~\ref{thm:theoremtwirling} is applicable also for general sets of free operations that are weaker than resource non-generating operations. These results may therefore be applicable to complementary scenarios. 
}%
\\

\textbf{\emph{Surpassing conventional limitations}.\,---} Distillation in the conventional sense is constrained by many no-go theorems that restrict what transformations can be achieved in certain regimes. 
Consider zero-error distillation ($\varepsilon=0$). In this case, conventional distillation protocols have significant limits: they cannot, for example, distil \emph{any} pure states from states which are highly mixed (full- or almost full-rank, depending on the theory)~\cite{PhysRevLett.125.060405,Regula2021fundamental,Fang2020no-go}, not even when many copies of input states are available, and not even probabilistically~\cite{PhysRevLett.125.060405,regula_2022}. Virtual distillation suffers from no such no-go limitation: even full-rank states allow for distillation with a finite overhead cost.
An even stronger limitation constrains the one-shot distillation from isotropic states $\rho_p$ in theories such as quantum entanglement, coherence, or magic. No free operation can improve the fidelity of $\rho_p$ with a maximally resourceful state, making distillation impossible from a single copy of $\rho_p$ for all small values of $\varepsilon$~\cite{kent_1998,regula_2021-4}; virtual distillation allows one to surpass such restrictions. 

Fig.~\ref{fig:results} illustrates this capability in contexts of coherence, entanglement, and magic. 
Here, we compare the virtual distillation rate with the conventional case for the three types of resources, demonstrating the lack of a noise threshold that makes the conventional rate diminish to zero, and a more continuous dependence of the virtual distillation rate on the smoothing parameter $\varepsilon$.  
In the case of coherence~\cite{aberg2006quantifying,Baumgratz14,RevModPhys.89.041003,regula2018one}, we employ maximally incoherent operations (MIOs) and dephasing-covariant incoherent operations (DIOs) and apply Theorem~\ref{prop:general_bounds_cost} to determine the overhead for MIOs and DIOs. 
For entanglement~\cite{PhysRevA.53.2046,PhysRevLett.78.2275,RevModPhys.81.865,Brandao2011oneshot}, we determine the overhead using Theorem~\ref{thm:theoremtwirling} for any arbitrary subset of separability-preserving (SEPP) operations that contains local operation and classical communication (LOCC). 
For magic~\cite{veitch2012negative,veitch2014resource,PhysRevLett.118.090501}, we examine stabilizer operations and obtain an exact formula for the overhead, as explained in the companion paper~\cite{sm}.

\vspace{0.2cm}
\textbf{\emph{Applications}.\,---}
 {We first consider an application in quantum teleportation, where entanglement is a critical resource.} 
Given noisy entangled states, conventional resource distillation generally assumes 
multi-qubit operations on many copies of the noisy state and the existence of quantum memory. 
Our virtual distillation protocol serves as a more experimentally-friendly alternative without these challenging requirements. For example, consider the noisy state
$\rho_{A'B}(p)=p\psi_{A'B} + (1-p)I_4/4$ with $\ket{\psi}_{A'B}=(\ket{00}_{A'B}+\ket{11}_{A'B})/\sqrt{2}$. For $p\ge 1/3$ (and assuming $\varepsilon=0$), we could virtually distill $\rho_{A'B}$ as 
        $\tilde \Lambda(\rho_{A'B}(p)) = \frac{4}{1+3p}\Lambda_+(\rho_{A'B}(p)) -\frac{3-3p}{1+3p} \Lambda_-(\rho_{A'B}(p))$,
    where 
$\Lambda_+$ is the identity channel and $\Lambda_-(\rho) =\sigma_{AB} = (I-\psi_{AB})/3$.
The distillation overhead is $C^0(\rho_{A'B}(p)) = \min\{(7-3p)/(1+3p),3\}$.
The cost remains a reasonable constant even if we consider virtual distillation to multiple noisy entangled states, a relevant case in distributed quantum computing~\cite{noteentanglement}.

Another setting to which our approach immediately applies is fault-tolerant quantum computation, where it is very natural to consider classical outcomes of quantum algorithms. It is then important to understand optimal ways to distill the underlying resource of magic~\cite{veitch2014resource,PhysRevLett.118.090501}. 
Multiple proposals have been made regarding a combination of quantum error mitigation and error correction methods~\cite{Lostaglio2021error,Piveteau2021error,Suzuki2021quantum}.
Our framework encompasses this strategy for magic state distillation as an application of the general approach of virtual resource distillation, in particular allowing us to extend the results in Ref.~\cite{Lostaglio2021error} to more general regimes~\cite{sm}.

\textbf{\emph{Discussion}.\,---}
We introduced virtual resource distillation, an extended experimentally friendly framework of resource distillation that integrates classical linear postprocessing into free operations. In doing so, we provided a fundamental tool for distilling the resources needed for quantum advantages in noisy current and near-future quantum technology. The results are general and applicable to many specific resource theories, such as coherence, entanglement, magic of quantum states, quantum communication, uniformity, athermality, as well as any other resource theory where the target state is pure. 
We also detail the much more general setting of virtual resource distillation of quantum channels and combs in the companion paper~\cite{sm}, enabling virtual distillation of non-Markovianity and quantum memory. This work considers discrete-variable resource theories, and an extension to continuous variables~\cite{PhysRevX.8.041038,PhysRevLett.126.110403} would also be interesting.

\vspace{0.2cm}
\begin{acknowledgments}
\noindent\textbf{Acknowledgments.\,---}
We thank Suguru Endo, Patrick Hayden, Jayne Thompson, and Mark M.\ Wilde for insightful discussions.
This work is supported by the National Natural Science Foundation of China Grant No.~12175003, the Agency for Science, Technology and Research (A*STAR) under its QEP2.0 programme (NRF2021-QEP2-02-P06), the Singapore Ministry of Education Tier 1 Grant RG77/22, the National Research Foundation, Singapore and the Singapore Ministry of Education Tier 2 Grant MOE-T2EP50221-0005. B.R.\ was partially supported by the Japan Society for the Promotion of Science (JSPS) KAKENHI Grant No.\ 22KF0067. R.T.\ was supported by the Lee Kuan Yew Postdoctoral Fellowship at Nanyang Technological University Singapore.

\end{acknowledgments}

\bibliographystyle{apsrmp4-2}
\bibliography{myref}

\begin{thebibliography}{54}%
\makeatletter
\providecommand \@ifxundefined [1]{%
 \@ifx{#1\undefined}
}%
\providecommand \@ifnum [1]{%
 \ifnum #1\expandafter \@firstoftwo
 \else \expandafter \@secondoftwo
 \fi
}%
\providecommand \@ifx [1]{%
 \ifx #1\expandafter \@firstoftwo
 \else \expandafter \@secondoftwo
 \fi
}%
\providecommand \natexlab [1]{#1}%
\providecommand \emph  [1]{``#1''}%
\providecommand \bibnamefont  [1]{#1}%
\providecommand \bibfnamefont [1]{#1}%
\providecommand \citenamefont [1]{#1}%
\providecommand \href@noop [0]{\@secondoftwo}%
\providecommand \href [0]{\begingroup \@sanitize@url \@href}%
\providecommand \@href[1]{\@@startlink{#1}\@@href}%
\providecommand \@@href[1]{\endgroup#1\@@endlink}%
\providecommand \@sanitize@url [0]{\catcode `\\12\catcode `\$12\catcode
  `\&12\catcode `\#12\catcode `\^12\catcode `\_12\catcode `\%12\relax}%
\providecommand \@@startlink[1]{}%
\providecommand \@@endlink[0]{}%
\providecommand \url  [0]{\begingroup\@sanitize@url \@url }%
\providecommand \@url [1]{\endgroup\@href {#1}{\urlprefix }}%
\providecommand \urlprefix  [0]{URL }%
\providecommand \Eprint [0]{\href }%
\providecommand \doibase [0]{http://dx.doi.org/}%
\providecommand \selectlanguage [0]{\@gobble}%
\providecommand \bibinfo  [0]{\@secondoftwo}%
\providecommand \bibfield  [0]{\@secondoftwo}%
\providecommand \translation [1]{[#1]}%
\providecommand \BibitemOpen [0]{}%
\providecommand \bibitemStop [0]{}%
\providecommand \bibitemNoStop [0]{.\EOS\space}%
\providecommand \EOS [0]{\spacefactor3000\relax}%
\providecommand \BibitemShut  [1]{\csname bibitem#1\endcsname}%
\let\auto@bib@innerbib\@empty
\bibitem [{\citenamefont {Horodecki}\ and\ \citenamefont
  {Oppenheim}(2013)}]{horodecki2013quantumness}%
  \BibitemOpen
  \bibfield  {author} {\bibinfo {author} {\bibfnamefont {M.}~\bibnamefont
  {Horodecki}}\ and\ \bibinfo {author} {\bibfnamefont {J.}~\bibnamefont
  {Oppenheim}},\ }\bibfield  {title} {\emph {\bibinfo {title} {(quantumness in
  the context of) resource theories},}\ }\href@noop {} {\bibfield  {journal}
  {\bibinfo  {journal} {Int. J. Mod. Phys. B}\ }\textbf {\bibinfo {volume}
  {27}},\ \bibinfo {pages} {1345019} (\bibinfo {year} {2013})}\BibitemShut
  {NoStop}%
\bibitem [{\citenamefont {Coecke}\ \emph {et~al.}(2016)\citenamefont {Coecke},
  \citenamefont {Fritz},\ and\ \citenamefont {Spekkens}}]{coecke_2016}%
  \BibitemOpen
  \bibfield  {author} {\bibinfo {author} {\bibfnamefont {B.}~\bibnamefont
  {Coecke}}, \bibinfo {author} {\bibfnamefont {T.}~\bibnamefont {Fritz}}, \
  and\ \bibinfo {author} {\bibfnamefont {R.~W.}\ \bibnamefont {Spekkens}},\
  }\bibfield  {title} {\emph {\bibinfo {title} {A mathematical theory of
  resources},}\ }\href {http://dx.doi.org/10.1016/j.ic.2016.02.008} {\bibfield
  {journal} {\bibinfo  {journal} {Inf. Comput.}\ }\textbf {\bibinfo {volume}
  {250}},\ \bibinfo {pages} {59} (\bibinfo {year} {2016})}\BibitemShut
  {NoStop}%
\bibitem [{\citenamefont {Chitambar}\ and\ \citenamefont
  {Gour}(2019)}]{chitambar2018quantum}%
  \BibitemOpen
  \bibfield  {author} {\bibinfo {author} {\bibfnamefont {E.}~\bibnamefont
  {Chitambar}}\ and\ \bibinfo {author} {\bibfnamefont {G.}~\bibnamefont
  {Gour}},\ }\bibfield  {title} {\emph {\bibinfo {title} {Quantum resource
  theories},}\ }\href {http://dx.doi.org/10.1103/RevModPhys.91.025001}
  {\bibfield  {journal} {\bibinfo  {journal} {Rev. Mod. Phys.}\ }\textbf
  {\bibinfo {volume} {91}},\ \bibinfo {pages} {025001} (\bibinfo {year}
  {2019})}\BibitemShut {NoStop}%
\bibitem [{\citenamefont {Bennett}\ \emph
  {et~al.}(1996{\natexlab{a}})\citenamefont {Bennett}, \citenamefont
  {Bernstein}, \citenamefont {Popescu},\ and\ \citenamefont
  {Schumacher}}]{PhysRevA.53.2046}%
  \BibitemOpen
  \bibfield  {author} {\bibinfo {author} {\bibfnamefont {C.~H.}\ \bibnamefont
  {Bennett}}, \bibinfo {author} {\bibfnamefont {H.~J.}\ \bibnamefont
  {Bernstein}}, \bibinfo {author} {\bibfnamefont {S.}~\bibnamefont {Popescu}},
  \ and\ \bibinfo {author} {\bibfnamefont {B.}~\bibnamefont {Schumacher}},\
  }\bibfield  {title} {\emph {\bibinfo {title} {Concentrating partial
  entanglement by local operations},}\ }\href
  {http://dx.doi.org/10.1103/PhysRevA.53.2046} {\bibfield  {journal} {\bibinfo
  {journal} {Phys. Rev. A}\ }\textbf {\bibinfo {volume} {53}},\ \bibinfo
  {pages} {2046} (\bibinfo {year} {1996}{\natexlab{a}})}\BibitemShut {NoStop}%
\bibitem [{\citenamefont {Vedral}\ \emph {et~al.}(1997)\citenamefont {Vedral},
  \citenamefont {Plenio}, \citenamefont {Rippin},\ and\ \citenamefont
  {Knight}}]{PhysRevLett.78.2275}%
  \BibitemOpen
  \bibfield  {author} {\bibinfo {author} {\bibfnamefont {V.}~\bibnamefont
  {Vedral}}, \bibinfo {author} {\bibfnamefont {M.~B.}\ \bibnamefont {Plenio}},
  \bibinfo {author} {\bibfnamefont {M.~A.}\ \bibnamefont {Rippin}}, \ and\
  \bibinfo {author} {\bibfnamefont {P.~L.}\ \bibnamefont {Knight}},\ }\bibfield
   {title} {\emph {\bibinfo {title} {Quantifying entanglement},}\ }\href
  {http://dx.doi.org/10.1103/PhysRevLett.78.2275} {\bibfield  {journal}
  {\bibinfo  {journal} {Phys. Rev. Lett.}\ }\textbf {\bibinfo {volume} {78}},\
  \bibinfo {pages} {2275} (\bibinfo {year} {1997})}\BibitemShut {NoStop}%
\bibitem [{\citenamefont {Horodecki}\ \emph {et~al.}(2009)\citenamefont
  {Horodecki}, \citenamefont {Horodecki}, \citenamefont {Horodecki},\ and\
  \citenamefont {Horodecki}}]{RevModPhys.81.865}%
  \BibitemOpen
  \bibfield  {author} {\bibinfo {author} {\bibfnamefont {R.}~\bibnamefont
  {Horodecki}}, \bibinfo {author} {\bibfnamefont {P.}~\bibnamefont
  {Horodecki}}, \bibinfo {author} {\bibfnamefont {M.}~\bibnamefont
  {Horodecki}}, \ and\ \bibinfo {author} {\bibfnamefont {K.}~\bibnamefont
  {Horodecki}},\ }\bibfield  {title} {\emph {\bibinfo {title} {Quantum
  entanglement},}\ }\href {http://dx.doi.org/10.1103/RevModPhys.81.865}
  {\bibfield  {journal} {\bibinfo  {journal} {Rev. Mod. Phys.}\ }\textbf
  {\bibinfo {volume} {81}},\ \bibinfo {pages} {865} (\bibinfo {year}
  {2009})}\BibitemShut {NoStop}%
\bibitem [{\citenamefont {Aberg}(2006)}]{aberg2006quantifying}%
  \BibitemOpen
  \bibfield  {author} {\bibinfo {author} {\bibfnamefont {J.}~\bibnamefont
  {Aberg}},\ }\bibfield  {title} {\emph {\bibinfo {title} {Quantifying
  superposition},}\ }\href@noop {} {\bibfield  {journal} {\bibinfo  {journal}
  {arXiv preprint
  \href{https://arxiv.org/abs/quant-ph/0612146}{quant-ph/0612146}}} (\bibinfo
  {year} {2006})}\BibitemShut {NoStop}%
\bibitem [{\citenamefont {Baumgratz}\ \emph {et~al.}(2014)\citenamefont
  {Baumgratz}, \citenamefont {Cramer},\ and\ \citenamefont
  {Plenio}}]{Baumgratz14}%
  \BibitemOpen
  \bibfield  {author} {\bibinfo {author} {\bibfnamefont {T.}~\bibnamefont
  {Baumgratz}}, \bibinfo {author} {\bibfnamefont {M.}~\bibnamefont {Cramer}}, \
  and\ \bibinfo {author} {\bibfnamefont {M.~B.}\ \bibnamefont {Plenio}},\
  }\bibfield  {title} {\emph {\bibinfo {title} {Quantifying coherence},}\
  }\href {http://dx.doi.org/10.1103/PhysRevLett.113.140401} {\bibfield
  {journal} {\bibinfo  {journal} {Phys. Rev. Lett.}\ }\textbf {\bibinfo
  {volume} {113}},\ \bibinfo {pages} {140401} (\bibinfo {year}
  {2014})}\BibitemShut {NoStop}%
\bibitem [{\citenamefont {Streltsov}\ \emph {et~al.}(2017)\citenamefont
  {Streltsov}, \citenamefont {Adesso},\ and\ \citenamefont
  {Plenio}}]{RevModPhys.89.041003}%
  \BibitemOpen
  \bibfield  {author} {\bibinfo {author} {\bibfnamefont {A.}~\bibnamefont
  {Streltsov}}, \bibinfo {author} {\bibfnamefont {G.}~\bibnamefont {Adesso}}, \
  and\ \bibinfo {author} {\bibfnamefont {M.~B.}\ \bibnamefont {Plenio}},\
  }\bibfield  {title} {\emph {\bibinfo {title} {Colloquium: Quantum coherence
  as a resource},}\ }\href {http://dx.doi.org/10.1103/RevModPhys.89.041003}
  {\bibfield  {journal} {\bibinfo  {journal} {Rev. Mod. Phys.}\ }\textbf
  {\bibinfo {volume} {89}},\ \bibinfo {pages} {041003} (\bibinfo {year}
  {2017})}\BibitemShut {NoStop}%
\bibitem [{\citenamefont {Veitch}\ \emph {et~al.}(2012)\citenamefont {Veitch},
  \citenamefont {Ferrie}, \citenamefont {Gross},\ and\ \citenamefont
  {Emerson}}]{veitch2012negative}%
  \BibitemOpen
  \bibfield  {author} {\bibinfo {author} {\bibfnamefont {V.}~\bibnamefont
  {Veitch}}, \bibinfo {author} {\bibfnamefont {C.}~\bibnamefont {Ferrie}},
  \bibinfo {author} {\bibfnamefont {D.}~\bibnamefont {Gross}}, \ and\ \bibinfo
  {author} {\bibfnamefont {J.}~\bibnamefont {Emerson}},\ }\bibfield  {title}
  {\emph {\bibinfo {title} {Negative quasi-probability as a resource for
  quantum computation},}\ }\href
  {http://dx.doi.org/10.1088/1367-2630/14/11/113011} {\bibfield  {journal}
  {\bibinfo  {journal} {New J. Phys.}\ }\textbf {\bibinfo {volume} {14}},\
  \bibinfo {pages} {113011} (\bibinfo {year} {2012})}\BibitemShut {NoStop}%
\bibitem [{\citenamefont {Veitch}\ \emph {et~al.}(2014)\citenamefont {Veitch},
  \citenamefont {Mousavian}, \citenamefont {Gottesman},\ and\ \citenamefont
  {Emerson}}]{veitch2014resource}%
  \BibitemOpen
  \bibfield  {author} {\bibinfo {author} {\bibfnamefont {V.}~\bibnamefont
  {Veitch}}, \bibinfo {author} {\bibfnamefont {S.~H.}\ \bibnamefont
  {Mousavian}}, \bibinfo {author} {\bibfnamefont {D.}~\bibnamefont
  {Gottesman}}, \ and\ \bibinfo {author} {\bibfnamefont {J.}~\bibnamefont
  {Emerson}},\ }\bibfield  {title} {\emph {\bibinfo {title} {The resource
  theory of stabilizer quantum computation},}\ }\href
  {http://dx.doi.org/10.1088/1367-2630/16/1/013009} {\bibfield  {journal}
  {\bibinfo  {journal} {New J. Phys.}\ }\textbf {\bibinfo {volume} {16}},\
  \bibinfo {pages} {013009} (\bibinfo {year} {2014})}\BibitemShut {NoStop}%
\bibitem [{\citenamefont {Howard}\ and\ \citenamefont
  {Campbell}(2017)}]{PhysRevLett.118.090501}%
  \BibitemOpen
  \bibfield  {author} {\bibinfo {author} {\bibfnamefont {M.}~\bibnamefont
  {Howard}}\ and\ \bibinfo {author} {\bibfnamefont {E.}~\bibnamefont
  {Campbell}},\ }\bibfield  {title} {\emph {\bibinfo {title} {Application of a
  resource theory for magic states to fault-tolerant quantum computing},}\
  }\href {http://dx.doi.org/10.1103/PhysRevLett.118.090501} {\bibfield
  {journal} {\bibinfo  {journal} {Phys. Rev. Lett.}\ }\textbf {\bibinfo
  {volume} {118}},\ \bibinfo {pages} {090501} (\bibinfo {year}
  {2017})}\BibitemShut {NoStop}%
\bibitem [{\citenamefont {Bennett}\ \emph
  {et~al.}(1996{\natexlab{b}})\citenamefont {Bennett}, \citenamefont
  {Bernstein}, \citenamefont {Popescu},\ and\ \citenamefont
  {Schumacher}}]{bennett_1996-1}%
  \BibitemOpen
  \bibfield  {author} {\bibinfo {author} {\bibfnamefont {C.~H.}\ \bibnamefont
  {Bennett}}, \bibinfo {author} {\bibfnamefont {H.~J.}\ \bibnamefont
  {Bernstein}}, \bibinfo {author} {\bibfnamefont {S.}~\bibnamefont {Popescu}},
  \ and\ \bibinfo {author} {\bibfnamefont {B.}~\bibnamefont {Schumacher}},\
  }\bibfield  {title} {\emph {\bibinfo {title} {Concentrating partial
  entanglement by local operations},}\ }\href
  {http://dx.doi.org/10.1103/PhysRevA.53.2046} {\bibfield  {journal} {\bibinfo
  {journal} {Phys. Rev. A}\ }\textbf {\bibinfo {volume} {53}},\ \bibinfo
  {pages} {2046} (\bibinfo {year} {1996}{\natexlab{b}})}\BibitemShut {NoStop}%
\bibitem [{\citenamefont {Kent}(1998)}]{kent_1998}%
  \BibitemOpen
  \bibfield  {author} {\bibinfo {author} {\bibfnamefont {A.}~\bibnamefont
  {Kent}},\ }\bibfield  {title} {\emph {\bibinfo {title} {Entangled {{Mixed
  States}} and {{Local Purification}}},}\ }\href
  {http://dx.doi.org/10.1103/PhysRevLett.81.2839} {\bibfield  {journal}
  {\bibinfo  {journal} {Phys. Rev. Lett.}\ }\textbf {\bibinfo {volume} {81}},\
  \bibinfo {pages} {2839} (\bibinfo {year} {1998})}\BibitemShut {NoStop}%
\bibitem [{\citenamefont {Rains}(2001)}]{rains2001semidefinite}%
  \BibitemOpen
  \bibfield  {author} {\bibinfo {author} {\bibfnamefont {E.~M.}\ \bibnamefont
  {Rains}},\ }\bibfield  {title} {\emph {\bibinfo {title} {A semidefinite
  program for distillable entanglement},}\ }\href
  {http://dx.doi.org/10.1109/18.959270} {\bibfield  {journal} {\bibinfo
  {journal} {IEEE Trans. Inf. Theory}\ }\textbf {\bibinfo {volume} {47}},\
  \bibinfo {pages} {2921} (\bibinfo {year} {2001})}\BibitemShut {NoStop}%
\bibitem [{\citenamefont {Brandao}\ and\ \citenamefont
  {Plenio}(2008)}]{brandao2008entanglement}%
  \BibitemOpen
  \bibfield  {author} {\bibinfo {author} {\bibfnamefont {F.~G.}\ \bibnamefont
  {Brandao}}\ and\ \bibinfo {author} {\bibfnamefont {M.~B.}\ \bibnamefont
  {Plenio}},\ }\bibfield  {title} {\emph {\bibinfo {title} {Entanglement theory
  and the second law of thermodynamics},}\ }\href
  {http://dx.doi.org/10.1038/nphys1100} {\bibfield  {journal} {\bibinfo
  {journal} {Nat. Phys.}\ }\textbf {\bibinfo {volume} {4}},\ \bibinfo {pages}
  {873} (\bibinfo {year} {2008})}\BibitemShut {NoStop}%
\bibitem [{\citenamefont {{Brandao}}\ and\ \citenamefont
  {{Datta}}(2011)}]{Brandao2011oneshot}%
  \BibitemOpen
  \bibfield  {author} {\bibinfo {author} {\bibfnamefont {F.~G. S.~L.}\
  \bibnamefont {{Brandao}}}\ and\ \bibinfo {author} {\bibfnamefont
  {N.}~\bibnamefont {{Datta}}},\ }\bibfield  {title} {\emph {\bibinfo {title}
  {One-shot rates for entanglement manipulation under non-entangling maps},}\
  }\href {http://dx.doi.org/10.1109/TIT.2011.2104531} {\bibfield  {journal}
  {\bibinfo  {journal} {IEEE Trans. Inf. Theory}\ }\textbf {\bibinfo {volume}
  {57}},\ \bibinfo {pages} {1754} (\bibinfo {year} {2011})}\BibitemShut
  {NoStop}%
\bibitem [{\citenamefont {Zhao}\ \emph {et~al.}(2018)\citenamefont {Zhao},
  \citenamefont {Liu}, \citenamefont {Yuan}, \citenamefont {Chitambar},\ and\
  \citenamefont {Ma}}]{zhao2018oneshot}%
  \BibitemOpen
  \bibfield  {author} {\bibinfo {author} {\bibfnamefont {Q.}~\bibnamefont
  {Zhao}}, \bibinfo {author} {\bibfnamefont {Y.}~\bibnamefont {Liu}}, \bibinfo
  {author} {\bibfnamefont {X.}~\bibnamefont {Yuan}}, \bibinfo {author}
  {\bibfnamefont {E.}~\bibnamefont {Chitambar}}, \ and\ \bibinfo {author}
  {\bibfnamefont {X.}~\bibnamefont {Ma}},\ }\bibfield  {title} {\emph {\bibinfo
  {title} {One-shot coherence dilution},}\ }\href
  {http://dx.doi.org/10.1103/PhysRevLett.120.070403} {\bibfield  {journal}
  {\bibinfo  {journal} {Phys. Rev. Lett.}\ }\textbf {\bibinfo {volume} {120}},\
  \bibinfo {pages} {070403} (\bibinfo {year} {2018})}\BibitemShut {NoStop}%
\bibitem [{\citenamefont {Regula}\ \emph {et~al.}(2018)\citenamefont {Regula},
  \citenamefont {Fang}, \citenamefont {Wang},\ and\ \citenamefont
  {Adesso}}]{regula2018one}%
  \BibitemOpen
  \bibfield  {author} {\bibinfo {author} {\bibfnamefont {B.}~\bibnamefont
  {Regula}}, \bibinfo {author} {\bibfnamefont {K.}~\bibnamefont {Fang}},
  \bibinfo {author} {\bibfnamefont {X.}~\bibnamefont {Wang}}, \ and\ \bibinfo
  {author} {\bibfnamefont {G.}~\bibnamefont {Adesso}},\ }\bibfield  {title}
  {\emph {\bibinfo {title} {One-shot coherence distillation},}\ }\href
  {http://dx.doi.org/10.1103/PhysRevLett.121.010401} {\bibfield  {journal}
  {\bibinfo  {journal} {Phys. Rev. Lett.}\ }\textbf {\bibinfo {volume} {121}},\
  \bibinfo {pages} {010401} (\bibinfo {year} {2018})}\BibitemShut {NoStop}%
\bibitem [{\citenamefont {Liu}\ \emph {et~al.}(2019)\citenamefont {Liu},
  \citenamefont {Bu},\ and\ \citenamefont {Takagi}}]{2019arXiv190405840L}%
  \BibitemOpen
  \bibfield  {author} {\bibinfo {author} {\bibfnamefont {Z.-W.}\ \bibnamefont
  {Liu}}, \bibinfo {author} {\bibfnamefont {K.}~\bibnamefont {Bu}}, \ and\
  \bibinfo {author} {\bibfnamefont {R.}~\bibnamefont {Takagi}},\ }\bibfield
  {title} {\emph {\bibinfo {title} {One-shot operational quantum resource
  theory},}\ }\href {http://dx.doi.org/10.1103/PhysRevLett.123.020401}
  {\bibfield  {journal} {\bibinfo  {journal} {Phys. Rev. Lett.}\ }\textbf
  {\bibinfo {volume} {123}},\ \bibinfo {pages} {020401} (\bibinfo {year}
  {2019})}\BibitemShut {NoStop}%
\bibitem [{\citenamefont {Regula}\ \emph {et~al.}(2020)\citenamefont {Regula},
  \citenamefont {Bu}, \citenamefont {Takagi},\ and\ \citenamefont
  {Liu}}]{Regula2020benchmarking}%
  \BibitemOpen
  \bibfield  {author} {\bibinfo {author} {\bibfnamefont {B.}~\bibnamefont
  {Regula}}, \bibinfo {author} {\bibfnamefont {K.}~\bibnamefont {Bu}}, \bibinfo
  {author} {\bibfnamefont {R.}~\bibnamefont {Takagi}}, \ and\ \bibinfo {author}
  {\bibfnamefont {Z.-W.}\ \bibnamefont {Liu}},\ }\bibfield  {title} {\emph
  {\bibinfo {title} {Benchmarking one-shot distillation in general quantum
  resource theories},}\ }\href {http://dx.doi.org/10.1103/PhysRevA.101.062315}
  {\bibfield  {journal} {\bibinfo  {journal} {Phys. Rev. A}\ }\textbf {\bibinfo
  {volume} {101}},\ \bibinfo {pages} {062315} (\bibinfo {year}
  {2020})}\BibitemShut {NoStop}%
\bibitem [{\citenamefont {Fang}\ and\ \citenamefont
  {Liu}(2020)}]{PhysRevLett.125.060405}%
  \BibitemOpen
  \bibfield  {author} {\bibinfo {author} {\bibfnamefont {K.}~\bibnamefont
  {Fang}}\ and\ \bibinfo {author} {\bibfnamefont {Z.-W.}\ \bibnamefont {Liu}},\
  }\bibfield  {title} {\emph {\bibinfo {title} {No-go theorems for quantum
  resource purification},}\ }\href
  {http://dx.doi.org/10.1103/PhysRevLett.125.060405} {\bibfield  {journal}
  {\bibinfo  {journal} {Phys. Rev. Lett.}\ }\textbf {\bibinfo {volume} {125}},\
  \bibinfo {pages} {060405} (\bibinfo {year} {2020})}\BibitemShut {NoStop}%
\bibitem [{\citenamefont {Vijayan}\ \emph {et~al.}(2020)\citenamefont
  {Vijayan}, \citenamefont {Chitambar},\ and\ \citenamefont
  {Hsieh}}]{Vijayan2020simple}%
  \BibitemOpen
  \bibfield  {author} {\bibinfo {author} {\bibfnamefont {M.~K.}\ \bibnamefont
  {Vijayan}}, \bibinfo {author} {\bibfnamefont {E.}~\bibnamefont {Chitambar}},
  \ and\ \bibinfo {author} {\bibfnamefont {M.-H.}\ \bibnamefont {Hsieh}},\
  }\bibfield  {title} {\emph {\bibinfo {title} {Simple bounds for one-shot
  pure-state distillation in general resource theories},}\ }\href
  {http://dx.doi.org/10.1103/PhysRevA.102.052403} {\bibfield  {journal}
  {\bibinfo  {journal} {Phys. Rev. A}\ }\textbf {\bibinfo {volume} {102}},\
  \bibinfo {pages} {052403} (\bibinfo {year} {2020})}\BibitemShut {NoStop}%
\bibitem [{\citenamefont {Regula}\ and\ \citenamefont
  {Takagi}(2021{\natexlab{a}})}]{Regula2021fundamental}%
  \BibitemOpen
  \bibfield  {author} {\bibinfo {author} {\bibfnamefont {B.}~\bibnamefont
  {Regula}}\ and\ \bibinfo {author} {\bibfnamefont {R.}~\bibnamefont
  {Takagi}},\ }\bibfield  {title} {\emph {\bibinfo {title} {Fundamental
  limitations on distillation of quantum channel resources},}\ }\href
  {http://dx.doi.org/https://doi.org/10.1038/s41467-021-24699-0} {\bibfield
  {journal} {\bibinfo  {journal} {Nat. Commun.}\ }\textbf {\bibinfo {volume}
  {12}},\ \bibinfo {pages} {4411} (\bibinfo {year}
  {2021}{\natexlab{a}})}\BibitemShut {NoStop}%
\bibitem [{\citenamefont {Fang}\ and\ \citenamefont
  {Liu}(2022)}]{Fang2020no-go}%
  \BibitemOpen
  \bibfield  {author} {\bibinfo {author} {\bibfnamefont {K.}~\bibnamefont
  {Fang}}\ and\ \bibinfo {author} {\bibfnamefont {Z.-W.}\ \bibnamefont {Liu}},\
  }\bibfield  {title} {\emph {\bibinfo {title} {No-go theorems for quantum
  resource purification: New approach and channel theory},}\ }\href
  {http://dx.doi.org/10.1103/PRXQuantum.3.010337} {\bibfield  {journal}
  {\bibinfo  {journal} {PRX Quantum}\ }\textbf {\bibinfo {volume} {3}},\
  \bibinfo {pages} {010337} (\bibinfo {year} {2022})}\BibitemShut {NoStop}%
\bibitem [{\citenamefont {Regula}\ and\ \citenamefont
  {Takagi}(2021{\natexlab{b}})}]{Regula2021oneshot}%
  \BibitemOpen
  \bibfield  {author} {\bibinfo {author} {\bibfnamefont {B.}~\bibnamefont
  {Regula}}\ and\ \bibinfo {author} {\bibfnamefont {R.}~\bibnamefont
  {Takagi}},\ }\bibfield  {title} {\emph {\bibinfo {title} {One-shot
  manipulation of dynamical quantum resources},}\ }\href
  {http://dx.doi.org/10.1103/PhysRevLett.127.060402} {\bibfield  {journal}
  {\bibinfo  {journal} {Phys. Rev. Lett.}\ }\textbf {\bibinfo {volume} {127}},\
  \bibinfo {pages} {060402} (\bibinfo {year} {2021}{\natexlab{b}})}\BibitemShut
  {NoStop}%
\bibitem [{\citenamefont {Takagi}\ \emph {et~al.}(2022)\citenamefont {Takagi},
  \citenamefont {Regula},\ and\ \citenamefont {Wilde}}]{Takagi2021oneshot}%
  \BibitemOpen
  \bibfield  {author} {\bibinfo {author} {\bibfnamefont {R.}~\bibnamefont
  {Takagi}}, \bibinfo {author} {\bibfnamefont {B.}~\bibnamefont {Regula}}, \
  and\ \bibinfo {author} {\bibfnamefont {M.~M.}\ \bibnamefont {Wilde}},\
  }\bibfield  {title} {\emph {\bibinfo {title} {One-shot yield-cost relations
  in general quantum resource theories},}\ }\href
  {http://dx.doi.org/10.1103/PRXQuantum.3.010348} {\bibfield  {journal}
  {\bibinfo  {journal} {PRX Quantum}\ }\textbf {\bibinfo {volume} {3}},\
  \bibinfo {pages} {010348} (\bibinfo {year} {2022})}\BibitemShut {NoStop}%
\bibitem [{\citenamefont {Regula}(2022{\natexlab{a}})}]{regula_2022}%
  \BibitemOpen
  \bibfield  {author} {\bibinfo {author} {\bibfnamefont {B.}~\bibnamefont
  {Regula}},\ }\bibfield  {title} {\emph {\bibinfo {title} {Probabilistic
  {{Transformations}} of {{Quantum Resources}}},}\ }\href
  {http://dx.doi.org/10.1103/PhysRevLett.128.110505} {\bibfield  {journal}
  {\bibinfo  {journal} {Phys. Rev. Lett.}\ }\textbf {\bibinfo {volume} {128}},\
  \bibinfo {pages} {110505} (\bibinfo {year} {2022}{\natexlab{a}})}\BibitemShut
  {NoStop}%
\bibitem [{\citenamefont {Takagi}\ \emph {et~al.}(2024)\citenamefont {Takagi},
  \citenamefont {Yuan}, \citenamefont {Regula},\ and\ \citenamefont {Gu}}]{sm}%
  \BibitemOpen
  \bibfield  {author} {\bibinfo {author} {\bibfnamefont {R.}~\bibnamefont
  {Takagi}}, \bibinfo {author} {\bibfnamefont {X.}~\bibnamefont {Yuan}},
  \bibinfo {author} {\bibfnamefont {B.}~\bibnamefont {Regula}}, \ and\ \bibinfo
  {author} {\bibfnamefont {M.}~\bibnamefont {Gu}},\ }\bibfield  {title} {\emph
  {\bibinfo {title} {Virtual quantum resource distillation: General framework
  and applications},}\ }\href {http://dx.doi.org/10.1103/PhysRevA.109.022403}
  {\bibfield  {journal} {\bibinfo  {journal} {Phys. Rev. A}\ }\textbf {\bibinfo
  {volume} {109}},\ \bibinfo {pages} {022403} (\bibinfo {year} {2024})},\
  \bibinfo {note} {companion paper}\BibitemShut {NoStop}%
\bibitem [{\citenamefont {Pashayan}\ \emph {et~al.}(2015)\citenamefont
  {Pashayan}, \citenamefont {Wallman},\ and\ \citenamefont
  {Bartlett}}]{PhysRevLett.115.070501}%
  \BibitemOpen
  \bibfield  {author} {\bibinfo {author} {\bibfnamefont {H.}~\bibnamefont
  {Pashayan}}, \bibinfo {author} {\bibfnamefont {J.~J.}\ \bibnamefont
  {Wallman}}, \ and\ \bibinfo {author} {\bibfnamefont {S.~D.}\ \bibnamefont
  {Bartlett}},\ }\bibfield  {title} {\emph {\bibinfo {title} {Estimating
  outcome probabilities of quantum circuits using quasiprobabilities},}\ }\href
  {http://dx.doi.org/10.1103/PhysRevLett.115.070501} {\bibfield  {journal}
  {\bibinfo  {journal} {Phys. Rev. Lett.}\ }\textbf {\bibinfo {volume} {115}},\
  \bibinfo {pages} {070501} (\bibinfo {year} {2015})}\BibitemShut {NoStop}%
\bibitem [{\citenamefont {Seddon}\ \emph {et~al.}(2021)\citenamefont {Seddon},
  \citenamefont {Regula}, \citenamefont {Pashayan}, \citenamefont {Ouyang},\
  and\ \citenamefont {Campbell}}]{seddon_2020}%
  \BibitemOpen
  \bibfield  {author} {\bibinfo {author} {\bibfnamefont {J.~R.}\ \bibnamefont
  {Seddon}}, \bibinfo {author} {\bibfnamefont {B.}~\bibnamefont {Regula}},
  \bibinfo {author} {\bibfnamefont {H.}~\bibnamefont {Pashayan}}, \bibinfo
  {author} {\bibfnamefont {Y.}~\bibnamefont {Ouyang}}, \ and\ \bibinfo {author}
  {\bibfnamefont {E.~T.}\ \bibnamefont {Campbell}},\ }\bibfield  {title} {\emph
  {\bibinfo {title} {Quantifying {{Quantum Speedups}}: {{Improved Classical
  Simulation From Tighter Magic Monotones}}},}\ }\href
  {http://dx.doi.org/10.1103/PRXQuantum.2.010345} {\bibfield  {journal}
  {\bibinfo  {journal} {PRX Quantum}\ }\textbf {\bibinfo {volume} {2}},\
  \bibinfo {pages} {010345} (\bibinfo {year} {2021})}\BibitemShut {NoStop}%
\bibitem [{\citenamefont {Yang}\ \emph {et~al.}(2021)\citenamefont {Yang},
  \citenamefont {Lu},\ and\ \citenamefont {Li}}]{PRXQuantum.2.040361}%
  \BibitemOpen
  \bibfield  {author} {\bibinfo {author} {\bibfnamefont {Y.}~\bibnamefont
  {Yang}}, \bibinfo {author} {\bibfnamefont {B.-N.}\ \bibnamefont {Lu}}, \ and\
  \bibinfo {author} {\bibfnamefont {Y.}~\bibnamefont {Li}},\ }\bibfield
  {title} {\emph {\bibinfo {title} {Accelerated quantum {M}onte {C}arlo with
  mitigated error on noisy quantum computer},}\ }\href
  {http://dx.doi.org/10.1103/PRXQuantum.2.040361} {\bibfield  {journal}
  {\bibinfo  {journal} {PRX Quantum}\ }\textbf {\bibinfo {volume} {2}},\
  \bibinfo {pages} {040361} (\bibinfo {year} {2021})}\BibitemShut {NoStop}%
\bibitem [{\citenamefont {Peng}\ \emph {et~al.}(2020)\citenamefont {Peng},
  \citenamefont {Harrow}, \citenamefont {Ozols},\ and\ \citenamefont
  {Wu}}]{PhysRevLett.125.150504}%
  \BibitemOpen
  \bibfield  {author} {\bibinfo {author} {\bibfnamefont {T.}~\bibnamefont
  {Peng}}, \bibinfo {author} {\bibfnamefont {A.~W.}\ \bibnamefont {Harrow}},
  \bibinfo {author} {\bibfnamefont {M.}~\bibnamefont {Ozols}}, \ and\ \bibinfo
  {author} {\bibfnamefont {X.}~\bibnamefont {Wu}},\ }\bibfield  {title} {\emph
  {\bibinfo {title} {Simulating large quantum circuits on a small quantum
  computer},}\ }\href {http://dx.doi.org/10.1103/PhysRevLett.125.150504}
  {\bibfield  {journal} {\bibinfo  {journal} {Phys. Rev. Lett.}\ }\textbf
  {\bibinfo {volume} {125}},\ \bibinfo {pages} {150504} (\bibinfo {year}
  {2020})}\BibitemShut {NoStop}%
\bibitem [{\citenamefont {{Brenner}}\ \emph {et~al.}(2023)\citenamefont
  {{Brenner}}, \citenamefont {{Piveteau}},\ and\ \citenamefont
  {{Sutter}}}]{Brenner2023optimal}%
  \BibitemOpen
  \bibfield  {author} {\bibinfo {author} {\bibfnamefont {L.}~\bibnamefont
  {{Brenner}}}, \bibinfo {author} {\bibfnamefont {C.}~\bibnamefont
  {{Piveteau}}}, \ and\ \bibinfo {author} {\bibfnamefont {D.}~\bibnamefont
  {{Sutter}}},\ }\bibfield  {title} {\emph {\bibinfo {title} {{Optimal wire
  cutting with classical communication}},}\ }\href@noop {} {\Eprint
  {http://arxiv.org/abs/2302.03366} {arXiv:2302.03366}  (\bibinfo {year}
  {2023})}\BibitemShut {NoStop}%
\bibitem [{\citenamefont {Li}\ and\ \citenamefont {Benjamin}(2017)}]{Li2017}%
  \BibitemOpen
  \bibfield  {author} {\bibinfo {author} {\bibfnamefont {Y.}~\bibnamefont
  {Li}}\ and\ \bibinfo {author} {\bibfnamefont {S.~C.}\ \bibnamefont
  {Benjamin}},\ }\bibfield  {title} {\emph {\bibinfo {title} {Efficient
  variational quantum simulator incorporating active error minimization},}\
  }\href {http://dx.doi.org/10.1103/PhysRevX.7.021050} {\bibfield  {journal}
  {\bibinfo  {journal} {Phys. Rev. X}\ }\textbf {\bibinfo {volume} {7}},\
  \bibinfo {pages} {021050} (\bibinfo {year} {2017})}\BibitemShut {NoStop}%
\bibitem [{\citenamefont {Endo}\ \emph {et~al.}(2021)\citenamefont {Endo},
  \citenamefont {Cai}, \citenamefont {Benjamin},\ and\ \citenamefont
  {Yuan}}]{doi:10.7566/JPSJ.90.032001}%
  \BibitemOpen
  \bibfield  {author} {\bibinfo {author} {\bibfnamefont {S.}~\bibnamefont
  {Endo}}, \bibinfo {author} {\bibfnamefont {Z.}~\bibnamefont {Cai}}, \bibinfo
  {author} {\bibfnamefont {S.~C.}\ \bibnamefont {Benjamin}}, \ and\ \bibinfo
  {author} {\bibfnamefont {X.}~\bibnamefont {Yuan}},\ }\bibfield  {title}
  {\emph {\bibinfo {title} {Hybrid quantum-classical algorithms and quantum
  error mitigation},}\ }\href {http://dx.doi.org/10.7566/JPSJ.90.032001}
  {\bibfield  {journal} {\bibinfo  {journal} {J. Phys. Soc. Jpn.}\ }\textbf
  {\bibinfo {volume} {90}},\ \bibinfo {pages} {032001} (\bibinfo {year}
  {2021})}\BibitemShut {NoStop}%
\bibitem [{\citenamefont {Temme}\ \emph {et~al.}(2017)\citenamefont {Temme},
  \citenamefont {Bravyi},\ and\ \citenamefont
  {Gambetta}}]{PhysRevLett.119.180509}%
  \BibitemOpen
  \bibfield  {author} {\bibinfo {author} {\bibfnamefont {K.}~\bibnamefont
  {Temme}}, \bibinfo {author} {\bibfnamefont {S.}~\bibnamefont {Bravyi}}, \
  and\ \bibinfo {author} {\bibfnamefont {J.~M.}\ \bibnamefont {Gambetta}},\
  }\bibfield  {title} {\emph {\bibinfo {title} {Error mitigation for
  short-depth quantum circuits},}\ }\href
  {http://dx.doi.org/10.1103/PhysRevLett.119.180509} {\bibfield  {journal}
  {\bibinfo  {journal} {Phys. Rev. Lett.}\ }\textbf {\bibinfo {volume} {119}},\
  \bibinfo {pages} {180509} (\bibinfo {year} {2017})}\BibitemShut {NoStop}%
\bibitem [{\citenamefont {{Buscemi}}\ \emph {et~al.}(2013)\citenamefont
  {{Buscemi}}, \citenamefont {{Dall'Arno}}, \citenamefont {{Ozawa}},\ and\
  \citenamefont {{Vedral}}}]{Buscemi2013twopoint}%
  \BibitemOpen
  \bibfield  {author} {\bibinfo {author} {\bibfnamefont {F.}~\bibnamefont
  {{Buscemi}}}, \bibinfo {author} {\bibfnamefont {M.}~\bibnamefont
  {{Dall'Arno}}}, \bibinfo {author} {\bibfnamefont {M.}~\bibnamefont
  {{Ozawa}}}, \ and\ \bibinfo {author} {\bibfnamefont {V.}~\bibnamefont
  {{Vedral}}},\ }\bibfield  {title} {\emph {\bibinfo {title} {{Direct
  observation of any two-point quantum correlation function}},}\ }\href@noop {}
  {\Eprint {http://arxiv.org/abs/1312.4240} {arXiv:1312.4240}  (\bibinfo {year}
  {2013})}\BibitemShut {NoStop}%
\bibitem [{\citenamefont {Jiang}\ \emph {et~al.}(2021)\citenamefont {Jiang},
  \citenamefont {Wang},\ and\ \citenamefont {Wang}}]{Jiang2021physical}%
  \BibitemOpen
  \bibfield  {author} {\bibinfo {author} {\bibfnamefont {J.}~\bibnamefont
  {Jiang}}, \bibinfo {author} {\bibfnamefont {K.}~\bibnamefont {Wang}}, \ and\
  \bibinfo {author} {\bibfnamefont {X.}~\bibnamefont {Wang}},\ }\bibfield
  {title} {\emph {\bibinfo {title} {Physical {I}mplementability of {L}inear
  {M}aps and {I}ts {A}pplication in {E}rror {M}itigation},}\ }\href
  {http://dx.doi.org/10.22331/q-2021-12-07-600} {\bibfield  {journal} {\bibinfo
   {journal} {{Quantum}}\ }\textbf {\bibinfo {volume} {5}},\ \bibinfo {pages}
  {600} (\bibinfo {year} {2021})}\BibitemShut {NoStop}%
\bibitem [{\citenamefont {Regula}\ \emph
  {et~al.}(2021{\natexlab{a}})\citenamefont {Regula}, \citenamefont {Takagi},\
  and\ \citenamefont {Gu}}]{Regula2021operational}%
  \BibitemOpen
  \bibfield  {author} {\bibinfo {author} {\bibfnamefont {B.}~\bibnamefont
  {Regula}}, \bibinfo {author} {\bibfnamefont {R.}~\bibnamefont {Takagi}}, \
  and\ \bibinfo {author} {\bibfnamefont {M.}~\bibnamefont {Gu}},\ }\bibfield
  {title} {\emph {\bibinfo {title} {Operational applications of the diamond
  norm and related measures in quantifying the non-physicality of quantum
  maps},}\ }\href {http://dx.doi.org/10.22331/q-2021-08-09-522} {\bibfield
  {journal} {\bibinfo  {journal} {{Quantum}}\ }\textbf {\bibinfo {volume}
  {5}},\ \bibinfo {pages} {522} (\bibinfo {year}
  {2021}{\natexlab{a}})}\BibitemShut {NoStop}%
\bibitem [{Note1()}]{Note1}%
  \BibitemOpen
  \bibinfo {note} {Specifically, according to the Hoeffding inequality~\cite
  {Hoeffding1963probability}, $\protect \mathcal {O}(\protect \qopname \relax
  o{log}(1/\delta )/\beta ^2)$ and $\protect \mathcal {O}(C^2\protect \qopname
  \relax o{log}(1/\delta )/\beta ^2)$ number of samples can estimate $\protect
  \tr [M\psi ^{\otimes m}]$ and $ \protect \tr [M\protect \tilde \Lambda (\rho
  )]$ to achieve an accuracy $\beta \ge 0$ with a failure probability less than
  $\delta \ge 0$.}\BibitemShut {Stop}%
\bibitem [{\citenamefont {Bravyi}\ and\ \citenamefont
  {Kitaev}(2005)}]{bravyi_2005}%
  \BibitemOpen
  \bibfield  {author} {\bibinfo {author} {\bibfnamefont {S.}~\bibnamefont
  {Bravyi}}\ and\ \bibinfo {author} {\bibfnamefont {A.}~\bibnamefont
  {Kitaev}},\ }\bibfield  {title} {\emph {\bibinfo {title} {Universal quantum
  computation with ideal {{Clifford}} gates and noisy ancillas},}\ }\href
  {http://dx.doi.org/10.1103/PhysRevA.71.022316} {\bibfield  {journal}
  {\bibinfo  {journal} {Phys. Rev. A}\ }\textbf {\bibinfo {volume} {71}},\
  \bibinfo {pages} {022316} (\bibinfo {year} {2005})}\BibitemShut {NoStop}%
\bibitem [{\citenamefont {Vandenberghe}\ and\ \citenamefont
  {Boyd}(1996)}]{sdp_1996}%
  \BibitemOpen
  \bibfield  {author} {\bibinfo {author} {\bibfnamefont {L.}~\bibnamefont
  {Vandenberghe}}\ and\ \bibinfo {author} {\bibfnamefont {S.}~\bibnamefont
  {Boyd}},\ }\bibfield  {title} {\emph {\bibinfo {title} {Semidefinite
  programming},}\ }\href {http://dx.doi.org/10.1137/1038003} {\bibfield
  {journal} {\bibinfo  {journal} {SIAM Review}\ }\textbf {\bibinfo {volume}
  {38}},\ \bibinfo {pages} {49} (\bibinfo {year} {1996})}\BibitemShut {NoStop}%
\bibitem [{\citenamefont {Brand{\~a}o}\ and\ \citenamefont
  {Plenio}(2010)}]{Brandao2010reversible}%
  \BibitemOpen
  \bibfield  {author} {\bibinfo {author} {\bibfnamefont {F.~G. S.~L.}\
  \bibnamefont {Brand{\~a}o}}\ and\ \bibinfo {author} {\bibfnamefont {M.~B.}\
  \bibnamefont {Plenio}},\ }\bibfield  {title} {\emph {\bibinfo {title} {A
  reversible theory of entanglement and its relation to the second law},}\
  }\href {http://dx.doi.org/10.1007/s00220-010-1003-1} {\bibfield  {journal}
  {\bibinfo  {journal} {Commun. Math. Phys.}\ }\textbf {\bibinfo {volume}
  {295}},\ \bibinfo {pages} {829} (\bibinfo {year} {2010})}\BibitemShut
  {NoStop}%
\bibitem [{\citenamefont {Regula}\ \emph {et~al.}(2019)\citenamefont {Regula},
  \citenamefont {Fang}, \citenamefont {Wang},\ and\ \citenamefont
  {Gu}}]{regula_2019-2}%
  \BibitemOpen
  \bibfield  {author} {\bibinfo {author} {\bibfnamefont {B.}~\bibnamefont
  {Regula}}, \bibinfo {author} {\bibfnamefont {K.}~\bibnamefont {Fang}},
  \bibinfo {author} {\bibfnamefont {X.}~\bibnamefont {Wang}}, \ and\ \bibinfo
  {author} {\bibfnamefont {M.}~\bibnamefont {Gu}},\ }\bibfield  {title} {\emph
  {\bibinfo {title} {One-shot entanglement distillation beyond local operations
  and classical communication},}\ }\href
  {http://dx.doi.org/10.1088/1367-2630/ab4732} {\bibfield  {journal} {\bibinfo
  {journal} {New J. Phys.}\ }\textbf {\bibinfo {volume} {21}},\ \bibinfo
  {pages} {103017} (\bibinfo {year} {2019})}\BibitemShut {NoStop}%
\bibitem [{\citenamefont {Zhao}\ \emph {et~al.}(2019)\citenamefont {Zhao},
  \citenamefont {Liu}, \citenamefont {Yuan}, \citenamefont {Chitambar},\ and\
  \citenamefont {Winter}}]{zhao2018oneshotdistill}%
  \BibitemOpen
  \bibfield  {author} {\bibinfo {author} {\bibfnamefont {Q.}~\bibnamefont
  {Zhao}}, \bibinfo {author} {\bibfnamefont {Y.}~\bibnamefont {Liu}}, \bibinfo
  {author} {\bibfnamefont {X.}~\bibnamefont {Yuan}}, \bibinfo {author}
  {\bibfnamefont {E.}~\bibnamefont {Chitambar}}, \ and\ \bibinfo {author}
  {\bibfnamefont {A.}~\bibnamefont {Winter}},\ }\bibfield  {title} {\emph
  {\bibinfo {title} {One-{{Shot Coherence Distillation}}: {{Towards
  Completing}} the {{Picture}}},}\ }\href
  {http://dx.doi.org/10.1109/TIT.2019.2911102} {\bibfield  {journal} {\bibinfo
  {journal} {IEEE Trans. Inf. Theory}\ }\textbf {\bibinfo {volume} {65}},\
  \bibinfo {pages} {6441} (\bibinfo {year} {2019})}\BibitemShut {NoStop}%
\bibitem [{\citenamefont {Regula}(2022{\natexlab{b}})}]{regula_2021-4}%
  \BibitemOpen
  \bibfield  {author} {\bibinfo {author} {\bibfnamefont {B.}~\bibnamefont
  {Regula}},\ }\bibfield  {title} {\emph {\bibinfo {title} {Tight constraints
  on probabilistic convertibility of quantum states},}\ }\href
  {http://dx.doi.org/10.22331/q-2022-09-22-817} {\bibfield  {journal} {\bibinfo
   {journal} {{Quantum}}\ }\textbf {\bibinfo {volume} {6}},\ \bibinfo {pages}
  {817} (\bibinfo {year} {2022}{\natexlab{b}})}\BibitemShut {NoStop}%
\bibitem [{not()}]{noteentanglement}%
  \BibitemOpen
  \href@noop {} {}\bibinfo {note} {When $\rho_{A'B}$ is a separable state, this
  represents the cost of classical simulation. An intuitive yet non-optimal
  simulation protocol would be using state tomography for Alice and sending the
  classical density matrix information to Bob. Our result instead gives an
  optimal simulation protocol. It is not hard to see that the overhead would
  increase 
  consider multi-round quantum teleportation between different parties, which
  reflects the necessity of using entangled states. For example, consider
  sequential teleportation from party 0 to party $n$, the overhead to simulate
  perfect teleportation using separable states would be $3^n$. On the other
  hand, suppose nearly entangled states
  $\rho(\varepsilon)=(1-\varepsilon)\psi_{A'B} + \varepsilon I_4/4$ are shared
  (with small $\varepsilon$), the overhead for simulating $n$ sequential rounds
  of teleportation is $[1+6\varepsilon/(4-3\varepsilon)]^n\approx
  e^{6n\varepsilon/(4-3\varepsilon)}$. Therefore, whenever
  $n\varepsilon=\mathcal O(1)$, the overhead is still a constant even for large
  $n$. For example, suppose $n=50$ and $\varepsilon=1\%$, the overhead is about
  $e^3\approx 20$.}\BibitemShut {Stop}%
\bibitem [{\citenamefont {Lostaglio}\ and\ \citenamefont
  {Ciani}(2021)}]{Lostaglio2021error}%
  \BibitemOpen
  \bibfield  {author} {\bibinfo {author} {\bibfnamefont {M.}~\bibnamefont
  {Lostaglio}}\ and\ \bibinfo {author} {\bibfnamefont {A.}~\bibnamefont
  {Ciani}},\ }\bibfield  {title} {\emph {\bibinfo {title} {Error mitigation and
  quantum-assisted simulation in the error corrected regime},}\ }\href
  {http://dx.doi.org/10.1103/PhysRevLett.127.200506} {\bibfield  {journal}
  {\bibinfo  {journal} {Phys. Rev. Lett.}\ }\textbf {\bibinfo {volume} {127}},\
  \bibinfo {pages} {200506} (\bibinfo {year} {2021})}\BibitemShut {NoStop}%
\bibitem [{\citenamefont {Piveteau}\ \emph {et~al.}(2021)\citenamefont
  {Piveteau}, \citenamefont {Sutter}, \citenamefont {Bravyi}, \citenamefont
  {Gambetta},\ and\ \citenamefont {Temme}}]{Piveteau2021error}%
  \BibitemOpen
  \bibfield  {author} {\bibinfo {author} {\bibfnamefont {C.}~\bibnamefont
  {Piveteau}}, \bibinfo {author} {\bibfnamefont {D.}~\bibnamefont {Sutter}},
  \bibinfo {author} {\bibfnamefont {S.}~\bibnamefont {Bravyi}}, \bibinfo
  {author} {\bibfnamefont {J.~M.}\ \bibnamefont {Gambetta}}, \ and\ \bibinfo
  {author} {\bibfnamefont {K.}~\bibnamefont {Temme}},\ }\bibfield  {title}
  {\emph {\bibinfo {title} {Error mitigation for universal gates on encoded
  qubits},}\ }\href {http://dx.doi.org/10.1103/PhysRevLett.127.200505}
  {\bibfield  {journal} {\bibinfo  {journal} {Phys. Rev. Lett.}\ }\textbf
  {\bibinfo {volume} {127}},\ \bibinfo {pages} {200505} (\bibinfo {year}
  {2021})}\BibitemShut {NoStop}%
\bibitem [{\citenamefont {Suzuki}\ \emph {et~al.}(2022)\citenamefont {Suzuki},
  \citenamefont {Endo}, \citenamefont {Fujii},\ and\ \citenamefont
  {Tokunaga}}]{Suzuki2021quantum}%
  \BibitemOpen
  \bibfield  {author} {\bibinfo {author} {\bibfnamefont {Y.}~\bibnamefont
  {Suzuki}}, \bibinfo {author} {\bibfnamefont {S.}~\bibnamefont {Endo}},
  \bibinfo {author} {\bibfnamefont {K.}~\bibnamefont {Fujii}}, \ and\ \bibinfo
  {author} {\bibfnamefont {Y.}~\bibnamefont {Tokunaga}},\ }\bibfield  {title}
  {\emph {\bibinfo {title} {Quantum error mitigation as a universal error
  reduction technique: Applications from the {NISQ} to the fault-tolerant
  quantum computing eras},}\ }\href
  {http://dx.doi.org/10.1103/PRXQuantum.3.010345} {\bibfield  {journal}
  {\bibinfo  {journal} {PRX Quantum}\ }\textbf {\bibinfo {volume} {3}},\
  \bibinfo {pages} {010345} (\bibinfo {year} {2022})}\BibitemShut {NoStop}%
\bibitem [{\citenamefont {Yadin}\ \emph {et~al.}(2018)\citenamefont {Yadin},
  \citenamefont {Binder}, \citenamefont {Thompson}, \citenamefont
  {Narasimhachar}, \citenamefont {Gu},\ and\ \citenamefont
  {Kim}}]{PhysRevX.8.041038}%
  \BibitemOpen
  \bibfield  {author} {\bibinfo {author} {\bibfnamefont {B.}~\bibnamefont
  {Yadin}}, \bibinfo {author} {\bibfnamefont {F.~C.}\ \bibnamefont {Binder}},
  \bibinfo {author} {\bibfnamefont {J.}~\bibnamefont {Thompson}}, \bibinfo
  {author} {\bibfnamefont {V.}~\bibnamefont {Narasimhachar}}, \bibinfo {author}
  {\bibfnamefont {M.}~\bibnamefont {Gu}}, \ and\ \bibinfo {author}
  {\bibfnamefont {M.~S.}\ \bibnamefont {Kim}},\ }\bibfield  {title} {\emph
  {\bibinfo {title} {Operational resource theory of continuous-variable
  nonclassicality},}\ }\href {http://dx.doi.org/10.1103/PhysRevX.8.041038}
  {\bibfield  {journal} {\bibinfo  {journal} {Phys. Rev. X}\ }\textbf {\bibinfo
  {volume} {8}},\ \bibinfo {pages} {041038} (\bibinfo {year}
  {2018})}\BibitemShut {NoStop}%
\bibitem [{\citenamefont {Regula}\ \emph
  {et~al.}(2021{\natexlab{b}})\citenamefont {Regula}, \citenamefont {Lami},
  \citenamefont {Ferrari},\ and\ \citenamefont
  {Takagi}}]{PhysRevLett.126.110403}%
  \BibitemOpen
  \bibfield  {author} {\bibinfo {author} {\bibfnamefont {B.}~\bibnamefont
  {Regula}}, \bibinfo {author} {\bibfnamefont {L.}~\bibnamefont {Lami}},
  \bibinfo {author} {\bibfnamefont {G.}~\bibnamefont {Ferrari}}, \ and\
  \bibinfo {author} {\bibfnamefont {R.}~\bibnamefont {Takagi}},\ }\bibfield
  {title} {\emph {\bibinfo {title} {Operational quantification of
  continuous-variable quantum resources},}\ }\href
  {http://dx.doi.org/10.1103/PhysRevLett.126.110403} {\bibfield  {journal}
  {\bibinfo  {journal} {Phys. Rev. Lett.}\ }\textbf {\bibinfo {volume} {126}},\
  \bibinfo {pages} {110403} (\bibinfo {year} {2021}{\natexlab{b}})}\BibitemShut
  {NoStop}%
\bibitem [{\citenamefont {Hoeffding}(1963)}]{Hoeffding1963probability}%
  \BibitemOpen
  \bibfield  {author} {\bibinfo {author} {\bibfnamefont {W.}~\bibnamefont
  {Hoeffding}},\ }\bibfield  {title} {\emph {\bibinfo {title} {Probability
  inequalities for sums of bounded random variables},}\ }\href
  {http://dx.doi.org/10.1080/01621459.1963.10500830} {\bibfield  {journal}
  {\bibinfo  {journal} {J. Am. Stat. Assoc.}\ }\textbf {\bibinfo {volume}
  {58}},\ \bibinfo {pages} {13} (\bibinfo {year} {1963})}\BibitemShut {NoStop}%
\end{thebibliography}%

\end{document}